\documentclass[prb,superscriptaddress,twocolumn,floatfix,amsmath,amssymb,showpacs]{revtex4}

\usepackage{graphicx}
\usepackage{dcolumn}
\usepackage{bm}

\begin{document}

\title{Insulator-to-metal transition and large thermoelectric effect in La$_{1-x}$Sr$_{x}$MnAsO}

\author{Yunlei Sun}
\affiliation{Department of Physics, Zhejiang University, Hangzhou
310027, China}

\author{Jinke Bao}
\affiliation{Department of Physics, Zhejiang University, Hangzhou
310027, China}

\author{Chunmu Feng}
\affiliation{Department of Physics, Zhejiang University, Hangzhou
310027, China}

\author{Zhu'an Xu}
\affiliation{Department of Physics, Zhejiang University, Hangzhou
310027, China} \affiliation{State Key Lab of Silicon Materials,
Zhejiang University, Hangzhou 310027, China}

\author{Guanghan Cao}
\email[corresponding author: ]{ghcao@zju.edu.cn}
\affiliation{Department of Physics, Zhejiang University, Hangzhou
310027, China} \affiliation{State Key Lab of Silicon Materials,
Zhejiang University, Hangzhou 310027, China}

\date{\today}

\begin{abstract}
We report the Sr substitution effect in an antiferromagnetic insulator LaMnAsO. The Sr doping limit is $x\sim$ 0.10 under the synthesis conditions, as revealed by x-ray diffractions indicate. Upon Sr doping, the room-temperature resistivity drops by five orders of magnitude down to $\sim$0.01 $\Omega\cdot$cm, and the temperature dependence of resistivity shows essentially metallic behavior for $x\geq$0.08. Hall and Seebeck measurements confirm consistently that the insulator-to-metal transition is due to hole doping. Strikingly, the room-temperature Seebeck coefficient for the metallic samples is as high as $\sim240 \mu$V/K, making the system as a possible candidate for thermoelectric applications.
\end{abstract}



\maketitle
\section{\label{sec:level1}Introduction}
The quaternary compounds with ZrCuSiAs-type (abbreviated as 1111-type) structure have recently attracted much interest, primarily because of the discovery of high temperature superconductivity in the iron arsenides.\cite{hosono,takahashi,cxh} The constituent elements in 1111 phase belong to different groups that are called hard acid, soft acid, soft base and hard base, respectively, according to the concept of "Hard and Soft Acids and Bases".\cite{HSAB} Therefore, the 1111 family holds many members owing to the combination of the four groups of elements. Up to 2008, there had been over 150 individuals synthesized.\cite{johrendt} Importantly, the element-selective feature in the four crystalline sites allows various kinds of successful chemical doping for inducing superconductivity in the prototype parent compound LaFeAsO.\cite{hosono,whh,Co,Ni,P}

The 1111 arsenides containing 3$d$ transition elements show diverse physical properties. LaMnAsO (LMAO) is an antiferromagnetic (AFM) semiconductor/insulator with pretty high Neel temperature of 317 K;\cite{emery1,emery2} LaFeAsO is an AFM\cite{AFM} semi-metal\cite{lzy}, serving as mother compound for Fe-based superconductors; LaCoAsO is an itinerant ferromagnet with Curie temperature of 66 K;\cite{hosono1} and LaNiAsO is a BCS-type superconductor with transition temperature $T_c\sim$3 K.\cite{ljl,hosono2} Fang and co-workers\cite{fz} performed first-principles calculations for the series of 1111 compounds. They were able to predict the AFM insulating ground state for LMAO, and the moment of Mn ions was calculated to be 3.1 $\mu_B$, consistent with the later experimental value of 3.34(2) $\mu_B$ at 2 K.\cite{emery2}

One notes that the properties of LMAO resembles those of the parent compound of cuprate superconductors. So, it is great interest to study its doping response. In fact, $Ln$Mn$Pn$O ($Ln$ stands for a lanthanide, $Pn$ refers to a pnictogen including P, As and Sb) was synthesized and structurally characterized over a decade ago.\cite{Mn1111} But it was only after the discovery of Fe-bsed superconductors that there has been growing research activities on this type of material.\cite{emery1,emery2,hosono3,hosono4,hosono5,aronson,bos,hiroi,tokura,1111Sb} The apparent valence of Mn in $Ln$Mn$Pn$O is 2+, thus the Mn ions have the electronic configuration of $d^5$, i.e., the valence electrons fill half the 3$d$ shell. Therefore, the Mn-based 1111 pnictides would be a Mott insulator, similar to the parent compound of cuprate superconductors, provided that the intrasite Coulomb correlation is high enough. Experiments basically agree with the point. Hosono and coworkers\cite{hosono3} showed that LaMn$Pn$O ($Pn$=P, As and Sb) thin films were all semiconducting with indirect bandgaps from 1.0 to 1.4 eV. They also demonstrated that LaMnPO was an AFM insulator with Neel temperature up to 375 K.\cite{hosono4} They found that the nominally undoped LaMnPO was an $n$-type semiconductor, and it could also be hole-doped with using Cu and Ca dopants, but no metallic conduction was achieved. Emery \emph{et. al.}\cite{emery1} reported that LMAO was a $C$-type AFM semiconductor with weak ferromagnetism as well as giant magnetoresistance. For $Ln$MnAsO containing magnetic rare-earth elements, there were several investigations on the magnetic orderings (and their interactions) of Mn spins and rare-earth moments,\cite{emery2,bos,hiroi} however, few doping study has been performed. Very recently Tokura and co-workers\cite{tokura} reported metallization in oxygen-deficient SmMnAsO$_{1-\delta}$ via electron doping. In this paper, we demonstrate an insulator-to-metal transition by hole doping in La$_{1-x}$Sr$_{x}$MnAsO. Surprisingly, the metallic sample has an unusually large Seebeck coefficient of $\sim240 \mu$V/K at room temperature (RT). This property seems to be attractive for thermoelectric applications.

\section{\label{sec:level2}Experimental}
Polycrystalline samples of La$_{1-x}$Sr$_{x}$MnAsO were synthesized
by solid state reaction in vacuum using powders of LaAs,
SrAs and MnO. LaAs and SrAs were presynthesized, respectively, by the reaction between La/Sr and As pieces at 1023-1123 K for 20 h. The purity of all the starting materials is no less than $\geq$ 99.9\%. The powders of LaAs,
SrAs and MnO were weighed according to the stoichiometric
ratios of La$_{1-x}$Sr$_{x}$MnAsO ($x$=0, 0.02, 0.04, 0.06, 0.08, 0.1 and 0.12), and thoroughly mixed in an agate mortar and pressed into pellets under a
pressure of 4000 kg/cm$^{2}$. The pellets were loaded in an alumina tube, and then sealed in a quartz-glass ampoule. We employed a glove box filled with protective argon atmosphere (the water and oxygen contents were below 0.1 ppm) to minimize the incorporation of excess oxygen. The ampoule was heated slowly to 1400 K in a muffle furnace, holding for 40 h and then cooled down to RT by switching off the furnace.

Powder x-ray diffraction (XRD) was carried out at RT on a D/Max-rA diffractometer with Cu-K$_{\alpha}$ radiation and a graphite monochromator. The XRD data were collected in a step-scan mode with a step of 0.02$^{\circ}$ (holding for 2 seconds) and scanning range of 10$^{\circ}$$\leq$2$\theta$$\leq$120$^{\circ}$. Lattice parameters were calculated by Rietveld refinements.

In the electrical transport measurement, thin (0.3 mm) bar-shaped specimen were cut and polished from the as-prepared pellets. Gold wires and silver paste were used for making the electrodes with the configurations suitable for the standard four-terminal resistivity ($\rho$) and Hall coefficient ($R_H$) measurements. The size of the contact pads leads to a total uncertainty in the absolute values of $\rho$ and $R_H$ of ¡À10 \%. The Seebeck coefficient ($S$) was measured by a conventional steady-state technique with a temperature gradient of $\sim$0.1 K/mm, on a Quantum Design physical property measurement system (PPMS-9). The magnetization measurements were performed on a Quantum Design magnetic property measurement system (MPMS-5).

\section{\label{sec:level3}Results and discussion}

Fig. 1(a) shows the XRD patterns of the as-prepared La$_{1-x}$Sr$_{x}$MnAsO samples.
Most of the peaks can be indexed by a tetragonal 1111-type unit cell with the space group $P4/nmm$. However, on close examining, several small peaks (with the intensity less than 5\% of that of the strongest peak) can be detected. Figure 1(b) shows a small peak close to the (200) reflections at 2$\theta=42.7^{\circ}$, which is ascribed to MnAs impurity. For the sample of $x$=0.12 (not shown here), unknown impurity phases grow substantially, suggesting that the doping limit is around  $x$=0.1. The lattice constants of the undoped LMAO are refined as $a$= 4.1202(5) \AA, and $c$=9.0467(10) \AA, which are basically consistent with those previously reported.\cite{emery1,Mn1111} With the Sr doping, the $c$ axis increases almost linearly up to $x$=0.1, meanwhile the $a$ axis hardly changes within the measurement errors [see figure 1(c)]. So, the cell volume increases with the Sr doping. This is reasonable, because the size of Sr$^{2+}$ is significantly larger than that of La$^{3+}$. The non-uniform variations in $a$ and $c$ probably reflect hole doping (see below), which ordinarily leads to shrinkage in the basal planes for layered oxide systems.

\begin{figure}
\includegraphics[width=8cm]{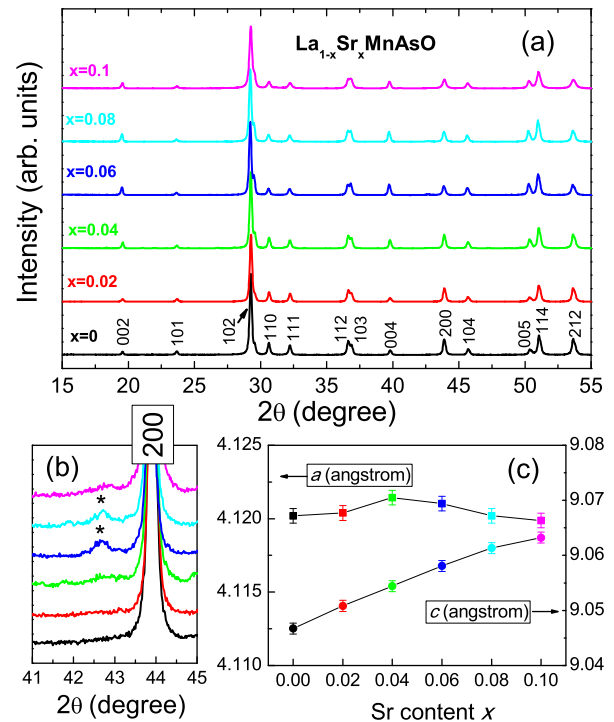}
\caption{(Color online) (a) Powder XRD patterns for the La$_{1-x}$Sr$_{x}$MnAsO samples, indexed by a tetragonal lattice with space group $P4/nmm$. (b) A magnified plot showing small amount of MnAs impurity for some samples. (c) Lattice constants as functions of Sr doping.}
\end{figure}

Fig. 2 shows the $\rho(T)$ data for the La$_{1-x}$Sr$_{x}$MnAsO polycrystalline samples. The undoped LMAO shows insulating behavior. The resistivity at RT is about 3000 $\Omega\cdot$cm, three times larger than the previously reported value\cite{emery1}. If fitting the $\rho(T)$ data near RT with the thermally activated conductivity, $\sigma=1/{\rho}\propto $ exp($-E_{a}$/$k_{B}T$), the activation energy will be 0.18 eV. This value is comparable to that (0.29 eV) in LaMnPO polycrystals.\cite{hosono4} In fact, the whole $\rho(T)$ data best fit the Mott's law of two-dimensional variable-range-hopping (VRH), ${\rho}\propto$ exp($T_{0}/T)^{1/3}$. This implies that a few charge carriers, which are localized, already exist in the undoped sample probably due to inevitable non-stoichiometry. Upon doping Sr, the resistivity decreases abruptly.  The RT resistivity decreases by over three orders of magnitude for the 2\% Sr doping, and the low-temperature $\rho(T)$ shows characteristic of weak localization. The sample of $x$=0.1 is essentially metallic, although the absolute resistivity is relatively large compared with conventional metals. The resistivity for $x$=0.08 sample has very weak temperature dependence, and it exhibits a minimum at $\sim$180 K, placing it on the boundary of insulator-metal transition. The resistivity upturn at low temperatures is probably due to weak localization and/or magnetic scattering. No other resistivity anomaly associated with the possible magnetic/electronic transition is observed for all the samples.

\begin{figure}
\includegraphics[width=8cm]{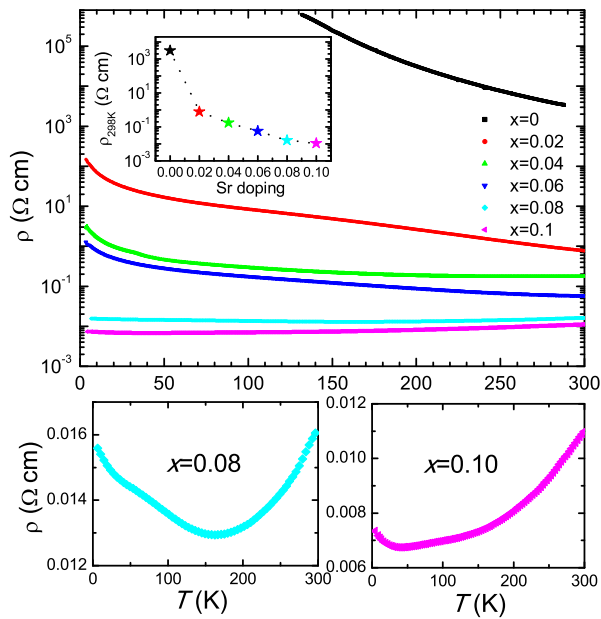}
\caption{(Color online) Temperature dependence of resistivity (with logarithmic scale) for La$_{1-x}$Sr$_{x}$MnAsO ploycrystalline samples. The lower panels show the $\rho(T)$ data of $x$=0.08 and 0.1, respectively, in ordinary scale.}
\end{figure}

It is noted that similar hole doping by Ca and Cu in LaMnPO failed to produce metallic conduction, although the RT resistivity was decreased by over three orders of magnitude.\cite{hosono4} The possible reason is that the solid-solution limit for Ca doping is smaller than that of Sr doping. This reminisces of the story that the hole-type superconductivity could be induced in LaFeAsO only by Sr doping (rather than Ca doping).\cite{whh} For the electron doping side, in comparison, an insulator-metal transition was not realized until $\delta$=0.2 in SmMnAsO$_{1-\delta}$.\cite{tokura} Therefore, the metallization in Mn-based 1111 material seems to be asymmetric to electron and hole doping. Hole doping more easily produces a metallic state.

In order to confirm and clarify the insulator-metal transition, we measured the samples' Hall coefficient. Fig. 3 shows the $R_H$ data for two representative samples: a semiconducting sample with $x$=0.02 and a metallic sample with $x$=0.1. Both samples show positive $R_H$ values in the whole temperature range, indicating that hole-type charge carriers dominate in the system. Assuming single band scenario, the charge carrier concentration can be easily estimated by $n_H=1/(eR_H$). So, the hole concentration for $x$=0.02 is 8.0$\times10^{24}$ m$^{-3}$ at RT, and it decreases rapidly with decreasing temperature. This behavior coincides with a heavily doped semiconductor. On the other hand, the estimated hole concentration at 10 K for $x$=0.1 is 1.5$\times10^{27}$ m$^{-3}$, and it changes hardly with temperature. The hole concentration corresponds to 0.11 holes/Mn, which meets the doping level of $x$=0.10 rather precisely. This result clearly indicates that the holes are doped via the Sr substitution in LMAO.

\begin{figure}
\includegraphics[width=7cm]{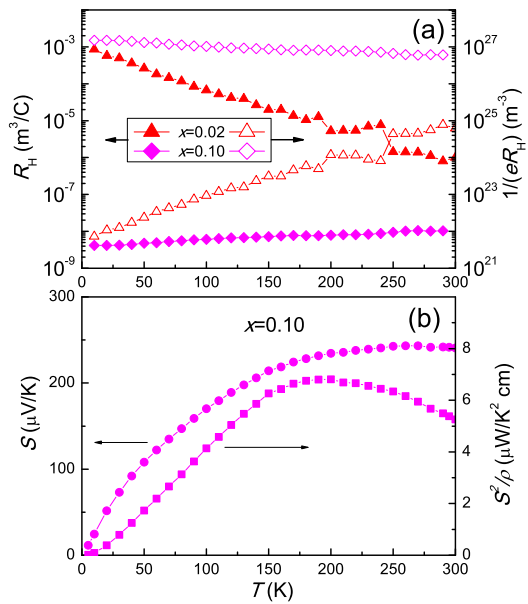}
\caption{(Color online) (a) Hall coefficient (left axis, solid symbols) and Hall carrier concentration (right axis, open symbols) of La$_{1-x}$Sr$_{x}$MnAsO ($x$=0.02 and 0.1) samples. (b)Seebeck coefficient (left axis) and power factor (right axis) for the $x$=0.1 sample.}
\end{figure}

The Seebeck coefficient ($S$) measured in La$_{1-x}$Sr$_{x}$MnAsO ($x$=0.02 to 0.1) are all positive, further confirming that the charge carriers are hole-type. Strikingly, the $S$ values are very large even for the metallic samples. Fig. 3(b) shows the $S(T)$ data for the sample of $x$=0.1. The thermopower tends to saturate near RT, achieving 240 $\mu$V/K, which is unusually large for a metal. It is significantly higher than that of a promising thermoelectric material Na$_x$CoO$_2$.\cite{terasaki} The reason for the high $S$ value is not clear, however, it could be associated with some special Fermi surface topology in LMAO. A recent theoretical calculation for a related compound BaMn$_2$As$_2$ (see below for further discussions) indeed shows that the Seebeck coefficient can be over 200 $\mu$V/K in the case of a small hole doping.\cite{band} The power factor, $S^{2}/\rho$, is over 5 $\mu$W/(K$^2$ cm) at RT. If the material is fabricated into the form of single crystals or thin films, whose resistivity could possibly decrease by one order of magnitude, the power factor may be over 50 $\mu$W/(K$^2$ cm), which is comparable to or even better than that of typical thermoelectric semiconductors, such as Bi$_2$Te$_3$.\cite{Bi2Te3}

Fig. 4 shows magnetic measurement data for the La$_{1-x}$Sr$_{x}$MnAsO samples. A weak ferromagnetic transition at $\sim$320 K is seen for all the samples. This result resembles the previous report of LMAO\cite{emery1} with a small saturated moment of 0.13 $\mu_B$/Mn. In comparison, our LMAO sample, which is "free" of MnAs impurity from XRD [see figure 1(b)], has even smaller saturated moment of 0.04 $\mu_B$/Mn. Is it intrinsic? We noted that the sample with the higher saturated magnetization contains more MnAs impurity. Since MnAs is a ferromagnet with Curie temperature at 318 K,\cite{MnAs} we conclude that the ferromagnetic signal comes from the MnAs, which has a saturated moment of 3.4 $\mu_B$/Mn.\cite{MnAs2} The concentration of MnAs impurity in our samples can thus be estimated to be from 1 at.\% to 6 at.\%. This explains why the MnAs impurity sometimes cannot be found by the XRD experiments.

\begin{figure}
\includegraphics[width=7cm]{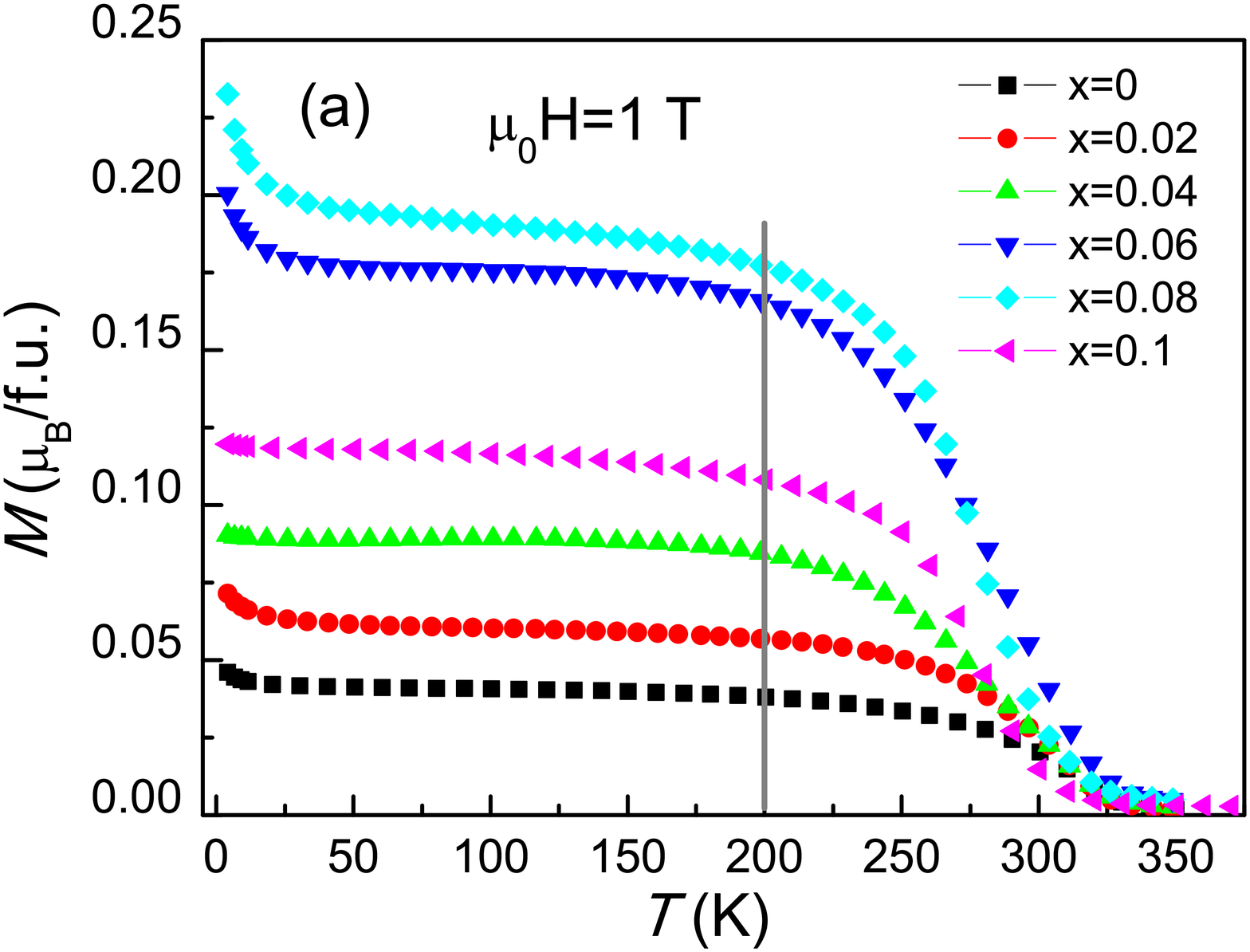}
\includegraphics[width=7cm]{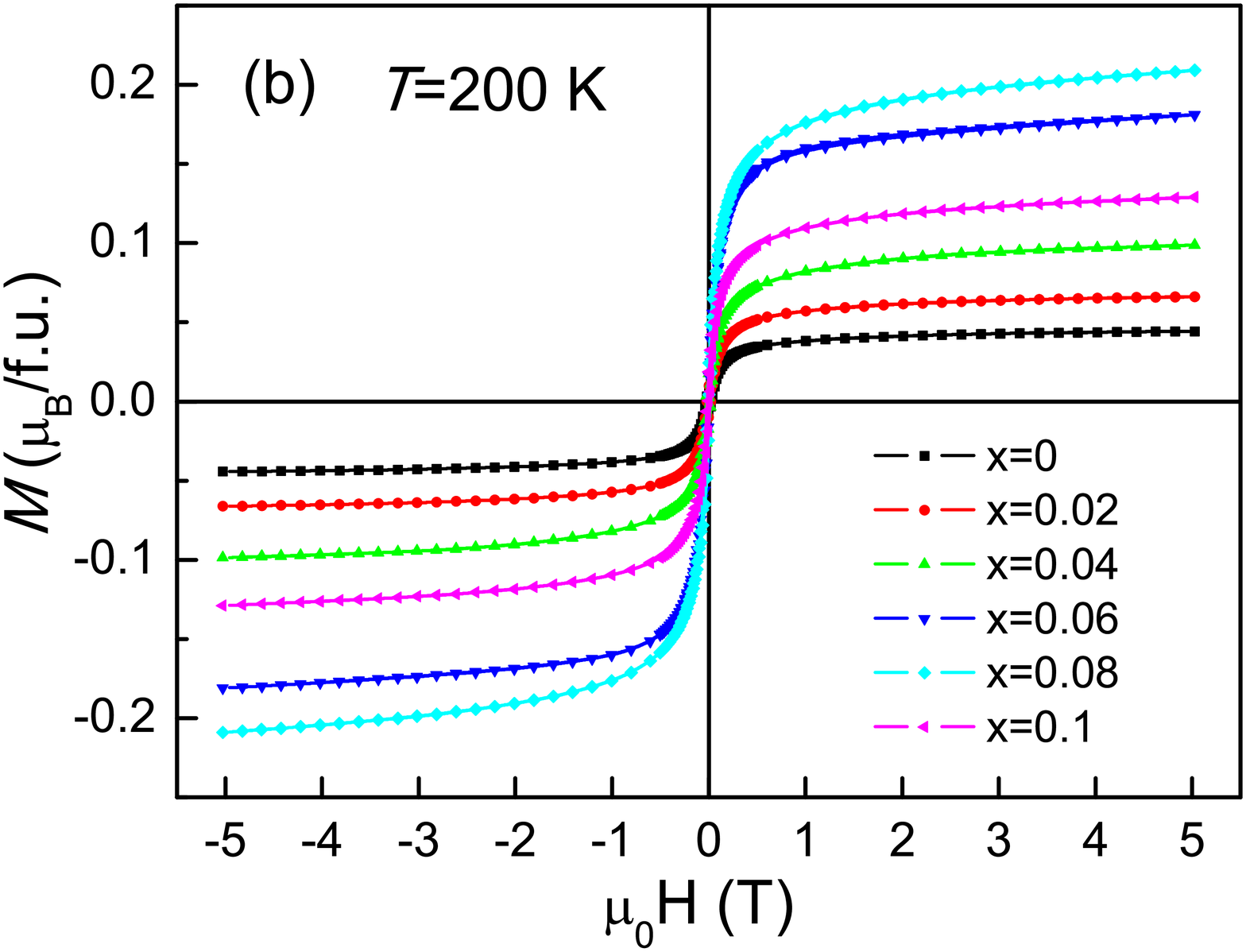}
\caption{(Color online) (a) Temperature dependence of magnetization for La$_{1-x}$Sr$_{x}$MnAsO. The ferromagnetic transition at 320 K is due to small amount of MnAs impurity. (b) Isothermal magnetization curves at 200 K for the same samples.}
\end{figure}

It is difficult to extract reliable quantitative information from the magnetization data which contain "huge" ferromagnetic background. Johnston and coworkers\cite{johnston1} succeeded to do so on studying BaMn$_2$As$_2$, because their crystal sample contained merely 0.11 at.\% MnAs impurity. Here we can only give some qualitative evaluations, based on the similar data analysis. First, after removing the saturated magnetization, the remaining paramagnetic signal tends to increase with increasing Sr doping. This enhanced paramagnetism should be contributed from the Pauli paramagnetism induced by hole doping. Second, the expected intrinsic AFM transitions was not detected, consistent with the absence of anomaly in above transport measurements. Future experiments of neutron diffractions and heat capacity measurements would be desirable to examine whether the Neel temperature and the Mn local-moment are suppressed or not by the Sr doping. Third, the low-temperature upturn in $M(T)$ curves seems to be extrinsic (due to unknown paramagnetic impurities and/or lattice imperfections), because such upturns are not reproducible for the sample with the identical composition.

Here we would like to compare LMAO with its close relative BaMn$_2$As$_2$, which contains the same MnAs-layers. BaMn$_2$As$_2$ was identified as a $G$-type AFM semiconductor with a Neel temperature as high as 625 K and an ordered moment of 3.88 $\mu_B$/Mn.\cite{johnston1,johnston2} The lowered moment (compared with a separate Mn$^{2+}$ ion) is ascribed to the substantial Mn3$d-$As4$p$ hybridizations by density functional calculations.\cite{band} Very recently, metallization in BaMn$_2$As$_2$ system was realized by hole doping with only 5\% K substitution.\cite{johnston3,note} The interesting result is that the local moment of Mn as well as the Neel temperature is basically preserved in the metallic doped samples. Under moderate pressures of $\sim$5 GPa, BaMn$_2$As$_2$ also turns into a metallic state.\cite{Mn-HP} Another recent report shows evidence of metallization by F doping as well as under high pressures.\cite{aronson} And it reveals that F doping in LaMnPO hardly changes the Neel temperature and the Mn local-moment.\cite{aronson} Similarly, the Sr doping probably does not suppress the magnetic order of Mn so much in LMAO. This suggests that in most Mn-based pnictides the doping-induced carriers dominantly have the As 4$p$ attribute, while the local moment of 3$d$ electrons is preserved.

\section{\label{sec:level4}Concluding remarks}

In summary, we have demonstrated an insulator-to-metal transition by hole doping in La$_{1-x}$Sr$_{x}$MnAsO, which is isostructural to the 1111 ferroarsenide high temperature superconductors. Electrical resistivity data indicate that the insulator-metal transition occurs at $x$=0.08. Hall coefficient measurement clearly shows that the holes are produced by the Sr doping. The metallic samples show unusually large Seebeck coefficient, which seems to be related to the Fermi surface topology. Thus the system is expected to be attractive for future thermoelectric applications.

The observed weak ferromagnetism is revealed to be due to an extrinsic effect (small amount of MnAs impurity). No anomalies associated with the AFM ordering were detected below 300 K, suggesting that the magnetism of Mn in LMAO is robust against the charge carrier doping. We propose that the doped holes mainly comes from the As 4$p$ orbitals. Further theoretical and experimental investigations are called for to clarify this issue.

\begin{acknowledgments}
This work is supported by the NSF of China (nos 90922002 and 10934005) and the National Basic Research Program of China (nos 2010CB923003 and 2011CBA00103).
\end{acknowledgments}

\end{document}